\documentstyle[preprint,aps,prb]{revtex}
\begin{document}
\draft
\title{Josephson vortex Cherenkov radiation}
\author{R.G.~Mints}
\address{School of Physics and Astronomy,\\ Raymond
and Beverly Sacler Faculty of Exact Sciences,\\ Tel Aviv
University,\\ Tel Aviv 69978, Israel}
\author{I.B.~Snapiro}
\address{Physics Department,\\ Technion--Israel
Institute of Technology,\\ Haifa 32000, Israel}
\maketitle
\begin{abstract}
We predict the Josephson vortex Cherenkov radiation of an
electromagnetic wave. We treat a long one--dimensional
Josephson junction. We consider the wave length of the
radiated electromagnetic wave to be much less than the
Josephson penetration depth. We use for calculations the
nonlocal Josephson electrodynamics. We find the expression
for the radiated power and for the radiation friction force
acting on a Josephson vortex and arising due to the
Cherenkov radiation. We calculate the relation between the
density of the bias current and the Josephson vortex
velocity.
\end{abstract}

\narrowtext
The Josephson vortex is a well--known and an important
example of a sin--Gordon soliton in solid state physics.
This soliton is a propagating non--linear wave describing
the phase difference between two weakly coupled
superconductors and the dynamics of the fluxons residing in
this contact \cite{1}. Results concerning the general features of
the motion of a Josephson vortex are interesting for
different systems in solid state physics where the sin--Gordon
soliton exists.
\par
Detailed knowledge of the Josephson vortex dynamics is
important for the flux dynamics and related phenomena in
superconductors, {\it e.g.}, flux creep, flux flow,
magnetization relaxation, current--voltage characteristics,
{\it etc}. Specific features of a Josephson vortex motion
are currently under thorough experimental and theoretical
study \cite{2,3,4,5}. In particular, very fast moving
Josephson vortices are observed and treated in annular
Josephson tunnel junctions \cite{2,3}.
\par
Josephson vortex dynamics is very important in
the novel layered high--temperature superconductors due to
their crystalline structure. In particular, the most
prominent $Bi$ and $Tl$ based copper oxide compounds consist
of a periodic stack of weakly coupled two--dimensional $CuO$
layers were the superconductivity presumably resides. In
this case a variety of linear crystalline structure defects
result from the crossing of the superconducting layers with
planar crystalline structure defects, {\it e.g.}, the grain
boundaries, twins, {\it etc}. These linear crystalline
structure defects can be treated as Josephson
junctions. The critical current density for the Josephson
junctions in the superconducting layers is relatively
high, especially for the coherent crystalline structure
defects, {\it e.g.}, for low--angle grain boundaries and
twins \cite{cur1,cur2}. The Josephson penetration length is
decreasing if the
Josephson critical current density is increasing. As a
result, the effect of the nonlocal Josephson electrodynamics
\cite{6,7} becomes important while treating the Josephson
vortices in the superconducting layers with the Josephson
penetration depth being of the order or less than the London
penetration depth.
\par
In this paper we consider the Josephson vortex Cherenkov
radiation of an electromagnetic wave. We treat a long
one--dimensional Josephson junction. We consider the wave
length of the radiated electromagnetic wave to be much less
than the Josephson penetration depth. We show that the
Josephson vortex velocity can be equal to the phase
velocity of this electromagnetic wave, which is the
necessary condition for the Cherenkov radiation. We use for
calculations the nonlocal Josephson electrodynamics. We
find the amplitude and power of the radiated wave and the
radiation friction force acting
on a Josephson vortex and arising due to the Cherenkov
radiation. We calculate the relation between the density of
the bias current across the Josephson junction and the
stationary Josephson vortex velocity. We consider the case
of a Josephson junction with a very high electrical
resistivity, {\it i.e.}, with a very low damping constant.
\par
The dynamics of a Josephson vortex in a one--dimensional
Josephson junction parallel to the $x$ axis is described
by the sine--Gordon equation for the space and time
dependent phase difference $\varphi (x,t)$. Taking into
account the damping resulting from the resistance of the
junction it reads
\begin{equation}
\varphi_{\tau\tau}-\varphi_{\xi\xi}+\eta\varphi_{\tau}+
\sin\varphi=\beta. \label{1}
\end{equation}
The subscripts $\tau$ and $\xi$ are to denote the
derivatives over the dimensionless time $\tau=t\omega_J$
and coordinate $\xi=x/\lambda_J$,
\begin{equation}
\omega_J=\sqrt{2ej_c\over \hbar C}\label{2}
\end{equation}
is the Josephson plasma frequency, $C$ is the specific
capacitance of the junction, $j_c$ is the critical current
density of the Josephson junction,
\begin{equation}
\lambda_J=\sqrt{c\Phi_0\over 16\pi^2\lambda j_c}
\label{3}
\end{equation}
is the Josephson penetration length, $\Phi_0$ is the flux
quantum, $\lambda$ is the London penetration depth,
\begin{equation}
\eta={1\over\omega_J RC}
\label{4}
\end{equation}
is the damping constant, $R$ is the specific resistance of
the junction, and $\beta=j/j_c$ is the dimensionless density
of the bias current across the junction.
\par
The well--known solution of Eq.~(\ref{1})
\begin{equation}
\varphi_0(x,t)=4\tan^{-1}
\exp\Bigl [{x-vt\over \lambda_J\sqrt{1-v^2/c^2_s}}\Bigr ],
\label{5}
\end{equation}
describes the uniform motion of a Josephson vortex with a
certain velocity $v$ in the case of zero dissipation and
zero driving force ($\gamma =\beta =0$). It follows from
Eq.~(\ref{5}) that in a long Josephson junction a Josephson vortex
moves similar to a relativistic particle with the highest
possible velocity $c_s=\lambda_J\omega_J$ (Swihart velocity)
\cite{1}.
\par
An electromagnetic wave with a specific dispersion relation
exists in a long one--dimensional Josephson
junction \cite{1}.
The solution of Eq.~(\ref{1}) in the form of a plain wave with a
small amplitude
\begin{equation}
\varphi (x,t)=\varphi_a \exp (-i\omega t+ikx), \qquad
\vert\varphi_a\vert\ll 1 \label{6}
\end{equation}
describes this electromagnetic wave.
\par
Let us consider the case of zero dissipation ($\eta =0$).
Then, the relation between the frequency $\omega$ and the
wave vector $k$ is given by the formula \cite{1}
\begin{equation}
\omega=\omega_J\ \sqrt{1+\lambda_J^2k^2},
\label{7}
\end{equation}
and the phase velocity of this electromagnetic wave
$v_\varphi$ is equal to
\begin{equation}
v_{\varphi}={\omega\over k}=
c_s\ \sqrt{1+{1\over k^2\lambda_J^2}}.
\label{8}
\end{equation}
\par
We can determine the phase difference $\varphi (x,t)$ in the
mainframe of local Josephson electrodynamics, {\it i.e.},
by the sin--Gordon equation, as long as
$\lambda\ll l_{\varphi}$, where $l_{\varphi}$ is the
characteristic space scale of $\varphi (x,t)$. It means, in
particular, that Eqs.~(\ref{5}), (\ref{7}) and (\ref{8})
are valid if
$\lambda\ll\lambda_J$ and $k\lambda\ll 1$. We show the
dependence $v_{\varphi}(k)$ calculated by means of
Eq.~(\ref{8})
by a solid line in Fig.~\ref{f1}.
\par
In the range of applicability of Eq.~(\ref{8}) the phase velocity
of the electromagnetic wave in a long Josephson junction is
a monotonically decreasing function of the wave vector.
It follows from Eq.~(\ref{8}) that for $k\lambda\ll 1$ the value
of $v_{\varphi}$ is higher than $c_s$. Since $c_s$ is the
highest possible velocity of a Josephson vortex it means
that there is no Josephson vortex Cherenkov radiation
\cite{8} of
an electromagnetic wave in the approximation of the local
Josephson electrodynamics.
\par

Let us now consider the general case, {\it i.e.}, the case
when the only restriction on the length scale $l_{\varphi}$
is given by the inequality $\xi\ll l_{\varphi}$, where
$\xi$ is the correlation length.
The relation between the phase difference $\varphi (x,t)$
and the magnetic field in the banks of the Josephson
junction is nonlocal if $l_{\varphi}<\lambda$. It results in
the nonlocal Josephson electrodynamics \cite{6,7}. In the general
form the equation generalizing Eq.~(\ref{1}) reads \cite{6}
\begin{equation}
\varphi_{\tau\tau}+\eta\varphi_{\tau}=
{\lambda_J^2\over\pi\lambda}\ \int\limits^\infty_{-\infty}
K_0\Bigl({|x-u|\over\lambda}\Bigr)\
{\partial^2\varphi\over\partial u^2}\ du -
\sin\varphi+\beta, \label{9}
\end{equation}
where $K_0(x)$ is the zero order modified Bessel function.
The integro--differential equation (\ref{9}) is valid as long as
$\xi\ll l_{\varphi}$.
\par
Using Eqs.~(\ref{6}) and (\ref{9}) we find the dispersion relation
$\omega (k)$ for an electromagnetic wave in a long Josephson
junction in the mainframe of the nonlocal Josephson
electrodynamics. In the case of zero dissipation ($\eta =0$)
it has the form
\begin{equation}
$$\omega=\omega_J\
\sqrt{1+{k^2\lambda_J^2\over\sqrt{1+k^2\lambda^2}}},
 \label{10}
\end{equation}
and thus the phase velocity of this electromagnetic wave is
equal to
\begin{equation}
v_{\varphi}={\omega\over k}
=c_s\ \sqrt{{1\over\sqrt{1+k^2\lambda^2}}+
{1\over k^2\lambda_J^2}}.
\label{11}
\end{equation}

Note, that the expressions given by Eqs.~(\ref{8}) and
(\ref{11}) for
$v_{\varphi}$ coincide if $k\lambda\ll 1$, {\it i.e.}, in
the range of validity of the local Josephson
electrodynamics.
\par
It follows from Eq.~(\ref{11}) that the electromagnetic
wave phase
velocity $v_{\varphi}$ given by Eq.~(\ref{11}) is a
monotonically
decreasing function of $k$. The value of $v_{\varphi}$
tends to zero when the wave vector $k$ tends to infinity. In
particular, in the limiting case $k\lambda\gg 1$ we have
\cite{5}:
\begin{equation}
v_{\varphi}\approx {c_s\over\sqrt{k\lambda}}\ll c_s,
\qquad k\lambda\gg 1.
\label{12}
\end{equation}
\par
We show the dependence $v_{\varphi}(k)$ given by
Eq.~(\ref{11})
by the dashed line in Fig.~\ref{f1}. We use for this plot
the value $\lambda_J=5\lambda$, {it i.e.},
$\lambda_J\gg\lambda$.
\par
Thus, there exist a certain region $k\ge k_c$, where the
phase velocity of an electromagnetic wave in a long
Josephson junction is lower than the highest possible
velocity of a Josephson vortex. We have the equation
$v_{\varphi}(k_c)=c_s$ to find the value of $k_c$. In case
when $\lambda\ll\lambda_J$ the solution of this equation is
given by an approximate formula
\begin{equation}
k_c\approx {1\over\lambda_J}\
\sqrt{\sqrt{2}{\lambda_J\over\lambda}+{3\over 4}}.
\label{13}
\end{equation}
\par
The existence of an electromagnetic wave with the phase
velocity lower than $c_s$
results in the Josephson vortex Cherenkov radiation. This
dissipation mechanism is especially effective when the
Josephson vortex velocity is approaching the highest
possible velocity $c_s$.
\par
The Josephson vortex Cherenkov radiation results, in
particular, in a
friction force acting on the radiating vortex. In order to
find this radiation friction force we solve the following
problem. Let us consider the uniform motion of a Josephson
vortex in a long Josephson junction. We treat the velocity
of this motion $v$ as a given constant value. We use for
calculations the perturbation theory, {\it i.e.}, we
neglect the dissipation arising due to the resistance of
the junction while considering the Josephson vortex
Cherenkov radiation of an electromagnetic wave.
\par
In order to find the amplitude of the radiated wave we look
for a solution of Eq.~(\ref{9}) with $\eta =0$ and $\beta =0$ in
the form
\begin{equation}
\varphi (x,t)=\varphi_0 (x-vt)+f(x,t),
\label{14}
\end{equation}
where $\varphi_0 (x-vt)$ is the phase difference given by
Eq.~(\ref{5}) for a single uniformly moving Josephson vortex and
$\vert f(x,t)\vert\ll 1$.
\par
A straightforward calculation shows that in the region
behind the front of the nonlinear wave $\varphi_0 (x-vt)$,
{\it i.e.}, in the region $(vt-x)\gg\lambda_J
\sqrt{1-v^2/c_s^2}$, the function $f(x,t)$ is a plain wave
taking the form
\begin{equation}
f(x,t)=f_0(k_p)\ \theta(vt-x)\exp[ik_p(x-vt)],
\label{15}
\end{equation}
where $\theta (x)$ is the $\theta$--function. The amplitude
of this plain wave $f_0(k_p)$ is given by the following
formula
\widetext
\begin{equation}
f_0(k_p)=\displaystyle\pi\ {v\over v_p-v}\
{k_p^2c_s^2-\omega_p^2+\omega_J^2\over\omega_p^2}
\,\displaystyle{1\over \cosh (0.5\pi k_p\lambda_J
\sqrt{1-v^2/c_s^2})},
\label{16}
\end{equation}
\narrowtext
where the wave vector $k_p$ is the root of the equation
\begin{equation}
\omega (k_p)=k_pv,
\label{17}
\end{equation}
the dispersion relation $\omega (k)$ is given by Eq.~(\ref{10}),
the frequency of the radiated wave $\omega_p=\omega (k_p)$,
and $v_p$ is the group velocity for a wave with $k=k_p$
\begin{equation}
v_p={\partial\omega\over\partial k}\Big\vert_{k_p}.
\label{18}
\end{equation}
\par
We explicitly use in the above calculations the phase
difference $\varphi_0(x,t)$ in the form given by Eq.~(\ref{5}).
It is valid as long as
$l_{\varphi}=\lambda_J\sqrt{1-v^2/c_s^2}\gg\lambda$, which
means that Eq.~(16) is valid if
\begin{equation}
\Bigl({\lambda\over\lambda_J}\Bigr)^2\ll
1-\Bigl({v\over c_s}\Bigr)^2.
\label{19}
\end{equation}
\par
The amplitude $f_0(k_p)$ is increasing when the velocity of
the Josephson vortex is approaching the highest possible
velocity $c_s$. In the vicinity of $c_s$, {\it i.e.}, in
the region $c_s-v\ll c_s$, Eq.~(\ref{16}) can be simplified
and
the values of $f_0(k_p)$, $k_p$ and $\omega_p$ can be found
in an explicit analytical form. The result of these
calculations is as follows
\begin{equation}
f_0(k_p)\approx 2\pi\exp\Bigl[-\pi\sqrt{2}\
{\lambda_J\over\lambda}\Bigl(1-{v\over c_s}\Bigr)\Bigr],
\qquad f_0(k_p)\ll 1
\label{20}
\end{equation}
\begin{equation}
k_p\approx{2\over\lambda}\sqrt{1-v/c_s},\qquad
{1\over\lambda_J}\ll k_p\ll {1\over\lambda},
\label{21}
\end{equation}
\begin{equation}
$$\omega_p\approx 2\omega_J\ {\lambda_J\over\lambda}
\sqrt{1-v/c_s},\qquad \omega_p\gg\omega_J.
\label{22}
\end{equation}
\par
Let us now find the radiation friction force $f_r$ acting
on a  unit length of a uniformly moving Josephson vortex.
To do it we use the energy conservation law equating $f_rv$
and $E_wv$, where $E_w$ is the electromagnetic wave energy
per unit area of the junction. It leads to the relation
$f_r=E_w$. Using the free energy functional corresponding
to the nonlocal Josephson electrodynamics \cite{6} we find that
\begin{equation}
E_w={\Phi_0^2\over 64\pi^3\lambda}\
{\omega_p^2\over c_s^2}\ f_0^2(k_p).
\label{23}
\end{equation}
\par
Thus, the Josephson vortex Cherenkov radiation of an
electromagnetic wave results in a radiation friction force
\begin{equation}
f_r={\Phi_0^2\over 4\pi\lambda^3}\
\Bigl(1-{v\over c_s}\Bigr)\
\exp\Bigl[-2\sqrt{2}\pi\ {\lambda_J\over\lambda}
\Bigl(1-{v\over c_s}\Bigr)\Bigr].
\label{24}
\end{equation}
The expression given by Eq.~(\ref{24}) is valid until the
absolute value of the exponent is bigger than one.
\par
Let us now consider a uniform motion of a Josephson vortex
in a Josephson junction with a bias current. In this case
the vortex is subjected to the Lorentz $f_L$ and the
radiation friction $f_r$ forces. The value of $f_L$ acting
per unit length of the vortex is equal to $\Phi_0j/c$.
Equating $f_L$ and $f_r$ we obtain the following relation
between the bias current density $j$ and the velocity of a
uniform motion of the Josephson vortex $v$
\begin{equation}
{j\over j_c}=4\pi\ {\lambda_J^2\over\lambda^2}\
\Bigl(1-{v\over c_s}\Bigr)\
\exp\Bigl[-2\sqrt{2}\pi\ {\lambda_J\over\lambda}
\Bigl(1-{v\over c_s}\Bigr)\Bigr].
\label{25}
\end{equation}
\par
A relation analogous to the one given by Eq.~(\ref{25}) and taking
into account only the damping due to the resistance of the
junction reads \cite{9}
\begin{equation}
{j\over j_c}={4\over\pi}\ \eta\ {v\over\sqrt{c_s^2-v^2}}.
\label{26}
\end{equation}
The dependence $j(v)$ given by Eq.~(26) is shown in Fig.~\ref{f2}
by the solid line. We use for this plot the value
$\eta=0.05$.
\par
Let us consider the Josephson vortex Cherenkov radiation
for $\eta\ne 0$. It follows from the energy conservation
law that for  $\eta\ll 1$ the dependence $j(v)$ is a sum of
the two dependencies given by Eqs.~(\ref{25}) and (\ref{26}),
{\it i.e.},
\widetext
\begin{equation}
{j\over j_c}=\displaystyle4\pi\
{\lambda_J^2\over\lambda^2}\
\Bigl(1-{v\over c_s}\Bigr)\
\exp\Bigl[-2\sqrt{2}\pi\ {\lambda_J\over\lambda}
\Bigl(1-{v\over c_s}\Bigr)\Bigr]+
\displaystyle{4\over\pi}\ \eta\ {v\over\sqrt{c_s^2-v^2}}.
\label{27}
\end{equation}
\narrowtext
\par
The dependence $j(v)$ given by Eq.~(\ref{27}) is shown in
Fig.~\ref{f2} by the dashed line. We use for this plot the
values $\eta=0.05$ and $\lambda_J=5\lambda$. It is seen
from Fig.~\ref{f2} that at $j\sim j_c$ the value of $v$ can
be significantly less than $c_s$.\par
The Josephson vortex velocity tends to a certain maximum,
$v_m$, when the current density tends to the critical
current density. Using Eq.~(\ref{27}) we can estimate $v_m$
as
\begin{equation}
1-{v_m\over c_s}\sim {1\over 2\sqrt{2}\pi}\,
{\lambda\over\lambda_J}.
\label{28}
\end{equation}
It follows from Eq.~(\ref{28}) that a noticeable difference
between $v_m$ and $c_s$ can be observed even if
$\lambda_J>\lambda$.\par
Note that, when the Josephson vortex velocity tends to its
maximum value $v_m$ the energy dissipation in the Josephson
junction is mainly due to the Josephson vortex Cherenkov
radiation, {\it i.e.}, the power release happens in the
form of electromagnetic radiation.\par
To summarize, we calculate the Josephson vortex Cherenkov
radiation of an electromagnetic wave in a long Josephson
junction. This dissipation mechanism results in the
radiation friction force and is especially effective if
the velocity of Josephson vortex is approaching the highest
possible velocity $c_s$. We find the relation between the
density of the bias current across the junction and the
Josephson vortex velocity.
\par
\
We are grateful to Dr. E.~Polturak for useful discussions.
This work was supported in part by the Foundation Raschi.

\begin{figure}
\caption{The dependence of the phase velocity
$v_\varphi$ on the wave vector $k$. The solid line
represents a plot using Eq.~(\protect \ref{8}), the dashed
line represents a plot using Eq.~(\protect \ref{11}).}
\label{f1}
\end{figure}

\begin{figure}
\caption{The dependence of the bias current
density $j$ on the Josephson vortex velocity $v$. The solid
line represents a plot using Eq.~(\protect \ref{26}), the
dashed line represents a plot using
Eq.~(\protect \ref{27}).}
\label{f2}
\end{figure}

\end{document}